\documentclass[a4paper]{jpconf}
\usepackage{graphicx}

\def\de{\partial}
\def\a{\alpha}
\def\b{\beta}
\def\g{\gamma}

\def\d{\delta}

{\rm }
\def\la{\lambda}
\def\La{\Lambda}
\def\k{\kappa}
\def\m{\mu}
\def\n{\nu}
\def\r{\rho}

\def\s{\sigma}

\def\th{\theta}
\def\z{\zeta}
\def\x{\chi}
\def\be{\begin{equation}}
 \def\ee{\end{equation}}
 \def\bea{\begin{eqnarray}}
 \def\eea{\end{eqnarray}}
 \def\a{\alpha}
 \def\b{\beta}
 \def\g{\gamma}
 \def\d{\delta}
 
 \def\s{\sigma}

\def\L{\Lambda}

\def\2{\frac{1}{2}}
\def\4{\frac{1}{4}}

\catcode`\@=11

\@addtoreset{equation}{section}

\def\@normalsize{\@setsize\normalsize{15pt}\xiipt\@xiipt
\abovedisplayskip 14pt plus3pt minus3pt%
\belowdisplayskip \abovedisplayskip
\abovedisplayshortskip  \z@ plus3pt%
\belowdisplayshortskip  7pt plus3.5pt minus0pt}
\def\small{\@setsize\small{13.6pt}\xipt\@xipt
\abovedisplayskip 13pt plus3pt minus3pt%
\belowdisplayskip \abovedisplayskip
\abovedisplayshortskip  \z@ plus3pt%
\belowdisplayshortskip  7pt plus3.5pt minus0pt
\def\@listi{\parsep 4.5pt plus 2pt minus 1pt
            \itemsep \parsep
            \topsep 9pt plus 3pt minus 3pt}}

\def\underline#1{\relax\ifmmode\@@underline#1\else
        $\@@underline{\hbox{#1}}$\relax\fi}
\@twosidetrue \relax

\catcode`@=12

\catcode`\@=11

\def\section{\@startsection{section}{1}{\z@}{3.5ex plus 1ex minus
   .2ex}{2.3ex plus .2ex}{\large\bf}}

\def\ps@headings{\def\@oddfoot{}\def\@evenfoot{}
\def\@oddhead{\hbox{}\hfill
        \makebox[.5\textwidth]{\raggedright\ignorespaces --\thepage{}--
        \hfill }}
\def\@evenhead{\@oddhead}
\def\subsectionmark##1{\markboth{##1}{}}
}

\ps@headings
\catcode`\@=12

\begin{document}
\title{Codimension-2 black hole solutions on a thin 3-brane 
and their extension into the bulk\footnote{To appear in the proceedings of the conference "NEB-XIII:Recent developments in gravity", held at Thessaloniki in June 2008.}}

\author{Minas~Tsoukalas}

\address{Department of Physics, National Technical University of
Athens, \\
Zografou Campus GR 157 73, Athens, Greece. \\}

\ead{minasts@central.ntua.gr}

\begin{abstract}
In this talk we discuss black hole solutions in six-dimensional gravity with a Gauss- 
Bonnet term in the bulk and an induced gravity term on a thin 3-brane of codimension-2. It is 
shown that these black holes can be localized on the 3-brane, and they can further be extended 
into the bulk by a warp function. These solutions have regular horizons and no other curvature 
singularities appear apart from the string-like ones. The projection of the Gauss-Bonnet term 
on the brane imposes a constraint relation which requires the presence of matter in the extra 
dimensions, in order to sustain our solutions.
\end{abstract}

\section{Introduction}

Taking a look at the last decade, a lot of attention has been drawn to brane-world models. These are mainly string inspired models \cite{Horava:1995qa}, where our universe is trapped on a brane, embedded in a higher dimensional space-time, the bulk. All matter fields are trapped on the brane and only gravity can propagate to the bulk. If we have n dimensions perpendicular to the brane, then we have a codimension-n brane-world \cite{ArkaniHamed:1998rs},\cite{RS12}. 

Of special interest, in the context of brane-worlds, is the study of black holes residing on the brane, mainly in the set up of  codimension-1 models. The issue was first addressed by Chamblin et.al. \cite{Chamblin:1999by}. A four-dimensional black hole on a 3-brane can be stretched to the extra dimension, forming something like a black cigar due to Gregory-Laflame instability \cite{Gregory:1993vy}, while at the same time solving the five-dimensional Einstein equations. Still  instabilities can occur \cite{BSINS}. Furthermore one can look into the effective four-dimensional Einstein equations on the brane \cite{Shiromizu:1999wj}, and try to solve them \cite{bbh}. Also recently there has been an investigation of brane-world black holes in the context of heterotic brane-world scenario \cite{Gregory:2009pu}. Investigating the case of a 2-brane in a four-dimensional universe there is a way out. Based on the form for an accelerating four-dimensional black hole (C-metric) \cite{Kinnersley:zw}
, Emparan et.al. \cite{EHM} found a BTZ black hole \cite{Banados:1992wn}  residing on a 2-brane, which can be extended into the bulk, forming  a BTZ black string. Their thermodynamic analysis showed that there is an instability similar to the Gregory-Laflamme instability \cite{Gregory:1993vy}.

Expanding the number of extra dimensions, one can investigate higher codimension models. The first attempt of course is to examine codimension-2 brane-worlds. Just like in three-dimensional gravity, where a point particle can induce a deficit angle, \cite{DJtH}, \cite{DJ}, the same can happen if we imbed a 3-brane into a six-dimensional space time. The vacuum energy of the brane, instead of curving the brane-wolrd volume, induces a deficit angle in the bulk solution around the brane \cite{Chen:2000at}. This property was the driving force, in order to solve the cosmological problem as was done by many authors \cite{6d}. Still due to near brane singularities, the brane's energy-momentum has to be proportional to its induced metric, if one wants to find nonsingular solutions \cite{Cline}. In order to avoid this obstacle one has to introduce some thickness to the brane \cite{Kanno:2004nr,Vinet:2004bk,Navarro:2004di} , or modify the gravitational action in the thin brane limit, introducing  either a
Gauss-Bonnet term~\cite{Bostock:2003cv} or an induced gravity term
on the brane~\cite{Papantonopoulos:2005ma}, recovering in this way four-dimensional gravity on the brane.

Unlike the situation in codimension-1 brane-world models, there is no clear understanding of time-dependent cosmological solutions in codimension-2 set ups. In the thin brane limit the brane equation of state and energy density are tuned, due to a relation between the energy-momentum tensors on the brane and in the bulk, giving this way a non-standard cosmology on the brane~\cite{Kofinas:2005py,Papantonopoulos:2005nw}. In order to resolve this problem one has
to regularize the codimension-2 branes by introducing some
thickness and then consider matter on them
\cite{regular,PST,ppz,tas}. To have a cosmological evolution on
the regularized branes the brane world-volume should be expanding
and in general
 the bulk space should also evolve in time. Alternatively ~\cite{Papantonopoulos:2007fk,Minamitsuji:2007fx} one has to consider a codimension-1 brane moving in the regularized
static background. The resulting cosmology, however, is unrealistic
having a negative Newton's constant (for a review on the cosmology
in six dimensions see~\cite{Papantonopoulos:2006uj}).

Deficit angles can also appear in the context of cosmic strings \cite{VilShel}.  For example, if a black hole is pierced by a cosmic string, a deficit angle appears in the space outside the string \cite{Aryal:1986sz,Achucarro:1995nu}. Taking advantage of that property, and adjusting it, a six-dimensional black hole localized on a 3-brane of codimension-2~\cite{Kaloper:2006ek} was constructed, with a conical structure in the bulk and deformations accommodating the deficit angle. Still due to near brane instabilities we mentioned earlier, it is not easy to realize these solutions in the thin brane limit, where high curvature terms are needed to accommodate matter on the brane. Finally application of rotation \cite{Kiley:2007wb} and perturbative analysis was performed~\cite{Chen:2007jz,alBinni:2007gk,Chen:2007ay,Rogatko:2008ut}.

In~\cite{CuadrosMelgar:2007jx} black
holes on an infinitely thin conical 2-brane and their extension
into a five-dimensional bulk with a Gauss-Bonnet term were studied. Two classes of solutions were found. The first class consists of the
familiar BTZ black hole which solves the junction conditions on a
conical 2-brane in vacuum. These solutions in the bulk are BTZ
string-like objects with regular horizons and no pathologies. The
warping to five-dimensions depends on the length $\sqrt{\alpha}$
where $\alpha$ is the Gauss-Bonnet coupling, and this length scale
defines the shape of the horizon. Consistency of the bulk
solutions requires a fine-tuned relation between the Gauss-Bonnet
coupling and the five-dimensional cosmological constant. The
second class of solutions consists of BTZ black holes with short
distance corrections. These solutions correspond to a BTZ black
hole conformally dressed with a scalar
field~\cite{Zanelli1996,Henneaux:2002wm}. Localization of these
black holes on the 2-brane leads to the interesting result that
the energy-momentum tensor required to support such solutions on
the brane corresponds to the energy-momentum tensor of a scalar
field in the
 limit $r/L_{3}<<1$, where $L_{3}$ is the length scale of the
three-dimensional AdS space and $r$ the radial distance on the
brane. Also these solutions have black string-like extensions into
the bulk.

In these notes we will present an extension of the work done in~\cite{CuadrosMelgar:2007jx}. Now insted of having a 2-brane in a five-dimensional space, we will have a 3-brane into a six-dimensional space-time~\cite{CuadrosMelgar:2008kn}. It is shown that  four-dimensional Schwarzschild-AdS black holes on the brane can have an extension into the bulk with a warp factor. These look like black string-like
objects with regular horizons. The warping to the extra dimensions, again depends on the Gauss-Bonnet coupling which is again fine-tuned to the six-dimensional cosmological constant. Keeping the deficit angle constant, the presence of a four dimensional black holes on the brane, requires matter in the two extra dimensions.

In the following, first we will present briefly the BTZ string-like solutions of the five-dimensional case. In section 3 we discuss the black holes on a 3-brane in a six-dimensional space-time and in section 4 we discuss the special r$\hat o$le
played by the Gauss-Bonnet term. Finally, in section 5 we
conclude.

\section{BTZ String-Like Solutions in Five-Dimensional Braneworlds of Codimension-2}

We consider the following gravitational action in five dimensions
with a Gauss-Bonnet term in the bulk and an induced
three-dimensional curvature term on the brane
\begin{eqnarray}\label{AcGBIG}
S_{\rm grav}&=&\frac{M^{3}_{5}}{2}\left\{ \int d^5 x\sqrt{-
g^{(5)}}\left[ R^{(5)}
+\alpha\left( R^{(5)2}-4 R^{(5)}_{MN}R^{(5)MN}+
R^{(5)}_{MNKL}R^{(5)MNKL}\right)\right] \right.\nonumber\\
&+& \left. r^{2}_{c} \int d^3x\sqrt{- g^{(3)}}\,R^{(3)}
\right\}+\int d^5 x \mathcal{L}_{bulk}+\int d^3 x
\mathcal{L}_{brane}\,,\label{5daction}
\end{eqnarray}
where $\alpha\, (\geq0)$ is the GB coupling constant and
$r_c=M_{3}/M_{5}^3$ is the induced gravity ``cross-over" scale
(marking the transition from 3D to 5D gravity).
We also consider the following bulk metric
\be ds_5^2=g_{\m\n}(x,\rho)dx^\m
dx^\n+a^{2}(x,\rho)d\rho^2+L^2(x,\rho)d\th^2~,\label{5dmetric} \ee
where $g_{\mu\nu}(x,0)$ is the brane-world metric and $x^{\mu}$
denote three  dimensions, $\mu=0,1,2$~ whereas $\rho,\th$ denote
the radial and angular coordinates of the two extra dimensions.
Capital $M$,~$N$ indices will take values in the five-dimensional
space. 

The Einstein equations resulting from the variation of the
action~(\ref{5daction}) are \be
 G^{(5)N}_M + r_c^2
G^{(3)\n}_\m g_M^\m g^N_\n {\d(\rho) \over 2 \pi L}-\alpha
H_{M}^{N} =\frac{1}{M^{3}_{5}} \left[T^{(B)N}_M+T^{(br)\n}_\m
g_M^\m g^N_\n {\d(\rho) \over 2 \pi L}\right]~, \label{einsequat3}
\ee where \bea
 H_M^N&=& \left[{1
\over 2}g_M^N (R^{(5)~2}
-4R^{(5)~2}_{KL}+R^{(5)~2}_{ABKL})\right.-2R^{(5)}R^{(5)N}_{M}\nonumber\\
&&+4R^{(5)}_{MP}R^{NP}_{(5)}\phantom{{1 \over 2}}~\left.
+4R^{(5)~~~N}_{KMP}R_{(5)}^{KP} -2R^{(5)}_{MKL P}R_{(5)}^{NKL
P}\right]~. \label{gaussbonnet} \eea To obtain the brane-world
equations we expand the metric around the brane as $
L(x,\rho)=\beta(x)\rho+O(\rho^{2})~. $ Function $L$ behaves as $L^{\prime}(x,0)=\beta(x)$, where a prime
denotes derivative with respect to $\rho$. We also demand that the
space in the vicinity of the conical singularity is regular which
imposes the supplementary conditions that $\de_\m \b=0$ and
$\partial_{\rho}g_{\mu\nu}(x,0)=0$~\cite{Bostock:2003cv}.

The extrinsic curvature is given by $K_{\m\n}=g'_{\m\n}$. Furthermore the second derivatives of the metric functions contain $\d$-function singularities at the position of
the brane \cite{Bostock:2003cv}. \bea
{L'' \over L}&=&-(1-L'){\d(\rho) \over L}+ {\rm non-singular~terms}~,\\
{K'_{\m\n} \over L}&=&K_{\m\n}{\d(\rho) \over L}+ {\rm
non-singular~terms}~. \eea

From the above singularity expressions and using the Gauss-Codazzi
equations, we can  match the singular parts of the Einstein
equations (\ref{einsequat3}) and get the following ``boundary"
Einstein equations
 \be G^{(3)}_{\m\n}={1 \over M_{(5)}^3 (r_c^2+8\pi
(1-\b)\a)}T^{(br)}_{\m\n}+{2\pi (1-\b) \over r_c^2+8\pi
(1-\b)\a}g_{\m\n} \label{einsteincomb3}~. \ee

We assume that there is a localized (2+1) black hole on the brane.
The brane metric is \be
ds_{3}^{2}=\left(-n(r)^{2}dt^{2}+n(r)^{-2}dr^{2}+r^{2}d\phi^{2}\right)~.
\label{3dmetric}\ee 
 We will look for black string solutions of the
 Einstein equations~(\ref{einsequat3}) using the
 five-dimensional metric~(\ref{5dmetric}) in the form
\be ds_5^2=f^{2}(\rho)\left(-n(r)^{2}dt^{2}+n(r)^{-2}dr^{2}+r^{2}
d\phi^{2}\right)+a^{2}(r,\rho)d\rho^2+L^2(r,\rho)d\th^2~.\label{5smetricc}
\ee

 The space outside the
 conical singularity is regular, therefore, we  demand that the warp function $ f(\rho) $ is
  also regular
 everywhere. We assume that there is only a cosmological constant $\Lambda_{5}$
in the bulk and we take $a(r,\rho)=1$. Then, from the bulk Einstein
equations \be G^{(5)}_{MN}-\alpha
H_{MN}=-\frac{\Lambda_{5}}{M^{3}_{5}}g_{MN}~,\ee combining the
$(rr,\phi \phi)$ equations we get

 \be
 \left(\dot{n}^{2}+n \ddot{n}-\frac{n \dot{n}}{r}\right)\left(1-4\alpha \frac{L''}{L}\right)=0~,\label{173}
 \ee
 while a combination of the $(\rho \rho, \theta \theta)$ equations
 gives
 \be
 \left(f''-\frac{f'L'}{L}\right)\left[3-4\frac{\alpha}{f^{2}}\left(\dot{n}^{2}+n
  \ddot{n}+2\frac{n \dot{n}}{r}+3f'^{2}
 \right)\right]=0\label{183}~,
 \ee
 where a  dot denotes derivatives with respect to $r$. The
 solutions of the equations (\ref{173}) and (\ref{183})
are summarized in the following table~\cite{CuadrosMelgar:2007jx}\\

\begin{table}
\caption{BTZ String-Like Solutions in Five-Dimensional Braneworlds
of Codimension-2}\label{table1}
\begin{center}
\begin{tabular}{lllllllll}
  \br
  $n(r)$ & $f(\r)$ & $L(\r)$ & $-\L_5$ & Constraints \\
  \mr
  BTZ & $\cosh\left(\frac{\r}{2\,\sqrt{\a}}\right)$ & $\forall L(\r)$ &
$\frac{3}{4\a}$ &
$L_3^2=4\,\a$ \\
  BTZ & $\cosh\left(\frac{\r}{2\,\sqrt{\a}}\right)$ & $2\,\b\,\sqrt{\a}\,\sinh\left(\frac{\r}{2\,\sqrt{\a}}\right)$ &
  $\frac{3}{4\a}$ & - \\
  BTZ & $\cosh\left(\frac{\r}{2\,\sqrt{\a}}\right)$ & $2\,\b\,\sqrt{\a}\,\sinh\left(\frac{\r}{2\,\sqrt{\a}}\right)$ &
  $\frac{3}{4\a}$ & $L_3^2=4\,\a$ \\
  BTZ & $\pm 1$ & $\frac{1}{\g}\,\sinh\left(\g\,\r\right)$ & $\frac{3}{l^2}$ & $\g=\sqrt{-\frac{2\L_5}{3+4\a\L_5}}$ \\
  $\forall n(r)$ & $\cosh\left(\frac{\r}{2\,\sqrt{\a}}\right)$ & $2\,\b\,\sqrt{\a}\,\sinh\left(\frac{\r}{2\,\sqrt{\a}}\right)$ &
  $\frac{3}{4\a}$ & -  \\
  $\sqrt{-M+\frac{r^2}{L_3^2}-\frac{\z}{r}}$ & $\cosh\left(\frac{\r}{2\,\sqrt{\a}}\right)$ &
$2\,\b\,\sqrt{\a}\,\sinh\left(\frac{\r}{2\,\sqrt{\a}}\right)$ &
$\frac{3}{4\a}$ &
$L_3^2=4\,\a$ \\
  $\sqrt{-M+\frac{r^2}{L_3^2}-\frac{\z}{r}}$ & $\pm 1$ & $2\,\b\,\sqrt{\a}\,\sinh\left(\frac{\r}{2\,\sqrt{\a}}\right)$
  & $\frac{1}{4\a}$ & $\L_5=-\frac{1}{4\a}=-\frac{3}{L_3^2}$ \\
  \br
\end{tabular}
\end{center}
\end{table}

In tab. 1, $L_3$ is the length scale of $AdS_{3}$ space.
Note that in all solutions there is a fine-tuned relation between
the Gauss-Bonnet coupling $\alpha$ and the five-dimensional
cosmological constant $\Lambda_{5}$, except for the solution in
the fourth row~\cite{CuadrosMelgar:2007jx}.

To introduce a brane we must solve the corresponding junction
conditions given by the Einstein equations on the brane
(\ref{einsteincomb3}) using the induced metric on the brane given
by (\ref{3dmetric}). In the case of a BTZ black hole $(n^{2}(r)=-M+\frac{r^{2}}{L_3^{2}}$), and a brane
cosmological constant given by $\Lambda_{3}=-1/L_3^{2}$, we
found that the energy-momentum tensor is null.

When $n(r)$ is of the form given by $
n(r)=\sqrt{-M+\frac{r^2}{L_3^2}-\frac{\z}{r}}~,
\label{nsolu5final} $ which is the BTZ black hole solution with
a short distance correction term, we find  that the matter source necessary to sustain such a solution
on the brane is given by $ T_\alpha ^\beta
=  \hbox{diag } \left(
\frac{\zeta}{2r^3},\frac{\zeta}{2r^3},-\frac{\zeta}{r^3} \right)\,
,\label{braneEnerMom} $ which is conserved on the
brane~\cite{Kofinas:2005a}. Interesting enough, for a scalar field
conformally coupled to BTZ~\cite{Zanelli1996,Henneaux:2002wm}, the
energy-momentum tensor needed to support such a solution at a
certain limit reduces to (\ref{braneEnerMom}) which is necessary
to localize this black hole on the conical 2-brane.

These solutions extend the brane BTZ black hole into the
bulk. Calculating the square of the Riemann tensor we find that at
the AdS horizon ($\rho \rightarrow \infty$) all solutions give
finite result and hence the only singularity is
  the  BTZ-corrected black hole singularity extended into the bulk.
  The warp function $f^{2}(\rho)$ gives the shape of a
'throat' to
    the horizon
of the BTZ string-like solution. The size of the horizon is
defined by the scale $\sqrt{\alpha}$ and this scale is fine-tuned
to the length scale of the five-dimensional AdS space.

\section{Black String-Like solutions in Six-Dimensional Braneworlds of Codimension-2 }

The
gravitational action similar to (\ref{AcGBIG}), but in six dimensions reads as
\begin{eqnarray}\label{act6}
S_{grav}&=& \frac{M_{6}^4}{2} \left\{ \int d^6 x \sqrt{-g^{(6)}} \left[ R^{(6)} + \alpha \left( R^{(6)2} - 4 R^{(6)}_{MN} R^{(6)MN} + R^{(6)}_{MNKL} R^{(6)MNKL} \right)\right] \right. \nonumber \\
&+&  \left. r_c^2 \int d^4 x \sqrt{-g^{(4)}} R^{(4)}
 \right\} + \int d^6 x {\cal L}_{bulk}
+ \int d^4 x {\cal L}_{brane}  \,.
\end{eqnarray}

 The metric as in the five-dimensional case is
\be ds_6^2=g_{\m\n}(r,\x)dx^\m
dx^\n+a^{2}(r,\x)d\x^2+L^2(r,\x)d\xi^2~,\label{6dmetric} \ee now
with $\mu=0,1,2,3$ whereas $\x,\xi$ denote the radial and angular
coordinates of the two extra dimensions (the $\x$ direction may
 or may not  be compact and the $\xi$ coordinate ranges form $0$ to $2\pi$).

The corresponding Einstein equations are
 \be
 G^{(6)N}_M + r_c^2
G^{(4)\n}_\m g_M^\m g^N_\n {\d(\x) \over 2 \pi L}-\alpha H_{M}^{N}
=\frac{1}{M^{4}_{6}} \left[-\L_{6}+T^{(B)N}_M+T^{(br)\n}_\m g_M^\m
g^N_\n {\d(\x) \over 2 \pi L}\right]~, \label{einsequat} \ee where
$H_{M}^{N}$ is the corresponding six-dimensional term of
(\ref{gaussbonnet})
 To obtain the braneworld equations we expand the
metric around the 3-brane as $ L(r,\x)=\b(r)\x+O(\x^{2})~, $
and as in the five-dimensional case the function $L$ behaves as
$L^{\prime}(r,0)=\beta(r)$, where a prime now denotes derivative
with respect to $\x$. The ``boundary" Einstein equations are \bea
G^{(4)}_{\m\n} \left(r_c^2+8\pi (1-\b)\a\right)|_0 &=& {1 \over
M_{6}^4}
T^{(br)}_{\m\n}|_0+ 2\pi (1-\b)g_{\m\n}|_0 \nonumber \\
&+& \pi L(r,\x)\,E_{\m\n}|_0-2\pi\b\a\,W_{\m\n}|_0
\label{einsteincomb}~, \eea where the term \be E_{\m\n}|_0 =
\left(K_{\m\n}-g_{\m\n}\,K\right)|_0~, \label{INDContrib} \ee
 appears because of the presence of the induced
gravity term in the gravitational action (we remind that $K_{\m\n}=g'_{\m\n}$), while the term \bea
W_{\m\n}|_0 &=&
g^{\la\s}\partial_{\x}g_{\m\la}\partial_{\x}g_{\n\s}|_0-
g^{\la\s}\partial_{\x}g_{\la\s}\partial_{\x}g_{\m\n}|_0 \nonumber \\
&+&\frac{1}{2}g_{\m\n}\left[\left(g^{\la\s}\partial_{\x}g_{\la\s}\right)^2
-g^{\la\s}g^{\d\r}\partial_{\x}g_{\la\d}\partial_{\x}g_{\s\r}\right]\Big{|}_0~,
\label{GBContrib} \eea is the Weyl term due to the presence of the
Gauss-Bonnet term in the bulk~\cite{Bostock:2003cv}.
  
If we demand that the space in the vicinity of the conical
singularity is regular ($\de_\m \b=0$) then (\ref{einsteincomb})
simply becomes~\cite{Bostock:2003cv,Papantonopoulos:2005ma} \be
G^{(4)}_{\m\n} \left(r_c^2+8\pi (1-\b)\a\right)|_0 = {1 \over
M_{6}^4} T^{(br)}_{\m\n}|_0+ 2\pi (1-\b)g_{\m\n}|_0
\label{einsteincombsimple}~. \ee

 We will look for black string solutions of the
 Einstein equations~(\ref{einsequat}) using the
following six-dimensional metric \be
ds_6^2=F^{2}(\x)\left(-A(r)^{2}dt^{2}+A(r)^{-2}dr^{2}+r^{2}d\th^{2}
+r^{2}\,\sin^2\th\,d\phi^{2}\right)+a^{2}(r,\x)d\x^2+L^2(r,\x)d\xi^2~.\label{6smetric}
\ee

Under the assumption that $T^{\x}_{\x}
= T^{\xi}_{\xi}$, combining the $rr - \th\th$ and $\x\x - \xi\xi$ components of the six-dimensional bulk equations we get similar expressions as in (\ref{173}), and (\ref{183}). The solutions are summarized in tab.  \ref{table2}.



\begin{table}
\caption{Black String-Like Solutions in Six-Dimensional
Braneworlds
of Codimension-2}\label{table2}
\begin{center}
\begin{tabular}{llllllll}
\hline 
$A^2(r)$ & $F(\chi)$ & $L(\chi)$ & $-\Lambda_6$ & Constraints \& $T^{(B)}$\\ 
\hline
$1+\frac{r^2}{L_4^2}-\frac{\zeta}{r}$ & $\cosh\left( \frac{\chi}{2\sqrt{3\alpha}}\right)$ & $\forall L(\chi)$  & $\frac{5}{12\a}$ & $\a=\frac{L_4^2}{12}$, \\
&&&&$T^{\chi}_\chi = T^{\xi}_\xi = - \frac{6\a \zeta^2}{r^6 F(\chi)^4}$\\
$1+\frac{r^2}{L_4^2}-\frac{\zeta}{r}$ & $\cosh\left( \frac{\chi}{2\sqrt{3\alpha}}\right)$ & $2\sqrt{3\alpha}\beta \sinh\left( \frac{\chi}{2\sqrt{3\alpha}}\right)$ & $\frac{5}{12\a}$ & $\a=\frac{L_4^2}{12}$, \\
&&&&$T^{\chi}_\chi = T^{\xi}_\xi = - \frac{6\a \zeta^2}{r^6 F(\chi)^4}$ \\
$1+\frac{r^2}{L_4^2}-\frac{\zeta}{r}$ & $\pm 1$ & $\frac{\b}{\g} \, \sinh{(\g\,\x)}$ & $\frac{6}{L_4^2} \left(1-\frac{2\a}{L_4^2}\right)$ & $\g = \frac{1}{L_4}\,\sqrt{\frac{1-\frac{L_4^2}{4\a}}{1-\frac{L_4^2}{12\a}}}$, \\
&&&&$T^{\chi}_\chi = T^{\xi}_\xi = - \frac{6\a \zeta^2}{r^6}$\\
$1+\frac{r^2}{L_4^2}-\frac{\zeta}{r}$ & $\pm 1$ & $\frac{\b}{\g} \chi \, \sinh{\g}$ & $\frac{6}{L_4^2} \left(1-\frac{2\a}{L_4^2}\right)$ & $\g = \frac{1}{L_4}\,\sqrt{\frac{1-\frac{L_4^2}{4\a}}{1-\frac{L_4^2}{12\a}}}$, \\
&&&&$T^{\chi}_\chi = T^{\xi}_\xi = - \frac{6\a \zeta^2}{r^6}$, \\
&&&&$T_t^{t}=T_r^{r}=T_\theta^{\theta}=T_\phi^{\phi}= \frac{3(4\a-L_4^2)}{L_4^2}$\\
(\ref{BH2?})& $\cosh \left(\frac{\x}{2\sqrt{3\alpha}}\right)$ & $2\,\sqrt{3\a}\,\b\, \sinh\left(\frac{\x}{2\sqrt{3\alpha}}\right)$ & $\frac{5}{12\a}$ & $\a=\frac{L_4^2}{12}$ \\
(\ref{nsolucte2}) &
$\pm 1$ & $2\sqrt{\a}\,\b \, \sinh{\left(\frac{\x}{2\sqrt{\a}}\right)}$ & $\frac{1}{4\a}$ & $\a=\frac{L_4^2}{4}$ \\
\br
\end{tabular}
\end{center}
\end{table}
Where
\be
A^{2}(r)=1+\frac{r^{2}}{L_4^2}\pm \sqrt{1 + \frac{C_3}{L_4^2} +
\frac{C_4}{L_4^4}\,r} ~,\label{BH2?} \ee 
and
\bea
A(r)^2&=&1+\frac{r^2}{4\a}-\frac{\sqrt{3}}{12\a}
\sqrt{2\,r^4 - 3\,C_4\,r +48\,\a\,\left(\a - C_3\right)}, \label{nsolucte2}
\eea

\section{The r$\hat{o}$le of the Gauss-Bonnet Term}

In codimension-2 brane-worlds there is a relation connecting the
Gauss-Bonnet term projected on the brane with the components of
the bulk energy-momentum tensor corresponding to the extra
dimensions~\cite{Papantonopoulos:2005ma}. In six dimensions it
reads \footnote{A similar relation involving the Gauss-Bonnet term was presented in ~\cite{Molina:2008kh} in a different context.}.
\be -\frac{1}{2}\,R^{(4)}|_0 - \frac{1}{2} \a  \left(
R^{(4)\,2} - 4R^{(4)\,2}_{\m\n} +R^{(4)\,2}_{\m\n\k\la}
\right)\Big{|}_0 = \frac{1}{M^4_6}\,T^{(B)\,\chi}_{\chi}|_0
-\frac{\La_6}{M^4_6}|_0\,. \label{6DBulkrr1} \ee All bulk
solutions have to satisfy this relation which acts as a
consistency relation. For the
Schwarzschild-AdS solution the square of
the Riemann tensor reads \be
 R_{\mu\nu\kappa\lambda}^2=\frac{192\zeta^{2}e^{\frac{4\chi}{L_4}}}{(1+e^{\frac{2\chi}{L_4}})^{4}r^{6}}
+\frac{60}{L_4^{4}}~,\label{riemanntensor} \ee while the Ricci
scalar and Ricci tensor are constants. Therefore, for the relation
(\ref{6DBulkrr1}) to be satisfied the bulk energy-momentum tensor
$T^{(B)\,\chi}_{\chi}|_0$ has to scale as $1/r^{6}$ with the right
coefficients. This is actually what happens considering the result shown in the table. Thus, the presence of the Gauss-Bonnet term in the
bulk, which acts as a source term because of its divergenceless
nature, dictates the form of matter that must be introduced in the
bulk in order to sustain a black hole on the brane\footnote{Black
hole solutions in codimension-2 braneworlds were also recently
discussed in~\cite{Charmousis:2008bt}.}. Investigating the nature of this matter we can see that at large $\chi$ goes to zero. Furthermore, on the brane, and at large distances we recover conventional four-dimensional gravity, while at small scales strong modifications appear.

\section{Conclusions}

The issue of localization of black holes in the context of brane-worlds is very difficult and still remains open. Here we discussed black holes residing on a thin brane of codimension-2 and their extension into the bulk, in the presence of a Gauss-Bonnet term in the bulk and an induced gravity term on the brane. Unlike the case of a 2-brane in a five-dimensional space-time filled with a cosmological constant, where a BTZ black hole can be smoothly extended to the bulk, the case of a 3-brane in a six-dimensional space-time is more tricky. Four-dimensional Schwarzschild-AdS black hole solutions can be extended into the bulk with a warp function, but now additional matter is needed to the transverse space in order to sustain the solution. 

The Gauss-Bonnet term alongside with the induced gravity term on the brane, which are needed in order to avoid pure tensional branes and to reproduce gravity on the brane, dictate the form of the matter which is needed. Still the nature of this matter remains undetermined. Furthermore there is the issue of stability. The Gauss-Bonnet term and the presence of matter make the task difficult to handle even in the case of the five-dimensional set up.

\section*{Acknowledgments}
The talk is based on work done in collaboration with B.Cuadros-Melgar, E. Papantonopoulos and V. Zamarias. We would like to thank the organizers of the 13th NEB Conference, held in Thessaloniki, for their kind hospitality and and perfect organization of the conference. This work was supported by the NTUA research program PEVE07.

\end{document}